\documentstyle[aps,prl,epsfig]{revtex}
\begin{document}
\title{Field Suppression of the Density-of-States: A Mechanism for Large Linear Magnetoresistance}
\draft
\author{D.P. Young, J.F. DiTusa, R.G. Goodrich, J. Anderson, S. Guo, and P.W. Adams}
\address{Department of Physics and Astronomy\\Louisiana State University\\Baton Rouge, Louisiana,
70803}
\author{Julia Y. Chan}
\address{Department of Chemistry\\Louisiana State University\\Baton Rouge, LA, 70803}
\author{Donavan Hall}
\address{National High Magnetic Field Laboratory\\Florida State University\\Tallahassee, FL, 32310}
\date{\today}
\maketitle
\begin{abstract}
Hall, resistivity, magnetization, and thermoelectric power measurements were performed on single crystals of the highly anisotropic
layered metal LaSb$_{2}$.  A 100-fold linear magnetoresistance (MR) was observed in fields up to 45 T, with no indication of
saturation. We show that the MR is associated with a magnetic-field-dependent holelike carrier density,
$n(H)\propto1/H$.  The effect is orbital, depending upon the component of the magnetic field normal to the layers.  At low
temperature, a field of 9 T reduces the carrier density by more than an order of magnitude.  
	\\     
\end{abstract}
\pacs{PACS numbers: 72.20.My, 72.15.Gd, 71.20.Lp}

	One of the most successful strategies for producing technologically relevant magnetoresistive materials is to enhance the effects
of field-dependent magnetic scattering processes through the  creation of magnetic superlattices \cite{GMR} or by doping magnetic
insulators such that a magnetic and metal-insulator transition coincide \cite{Manganite}.  Unexpectedly, several recent discoveries
of a large magnetoresistance (MR) in low carrier density $\em non$-magnetic metals \cite{Lifshits,Falicov,McClure,Adams} and
semiconductors \cite{Rosenbaum} suggest that there may be other ways of realizing field sensitive materials. 
However, progress to this end has been hampered by the lack of a concise understanding of the physical mechanisms that produce the
large linear positive MR found in Ag$_{2+\delta}$Te and Ag$_{2+\delta}$Se \cite{Rosenbaum}, LaSb$_2$ \cite{Budko}, as well as
several other non-magnetic materials \cite{Abrikosov1}.  In this Letter we show that the large linear MR in LaSb$_2$ is a
consequence of an orbital field suppression of the holelike carrier density, $n$.  The field suppression of $n$ accounts for a
linear MR that changes by a factor of 90 by 45 T, with no signs of saturation, suggesting that the system may undergo a
field-induced metal-insulator transition at higher fields.  The characterization of this density mediated MR may be an important
step towards understanding how electronic structure, dimensionality, disorder, and/or correlations give rise to similar linear MR
effects in the magnetic rare-earth diantimonide series $R$Sb$_2$, and perhaps, other magnetoresistive materials as well.

	LaSb$_2$ is a member of the $R$Sb$_{2}$ ($R$=La-Nd, Sm) family of compounds that all form in the orthorhombic SmSb$_{2}$ structure
\cite{Wang,Hullinger}. This is a highly anisotropic layered structure in which alternating La/Sb layers and two-dimensional
rectangular sheets of Sb atoms are stacked along the $c-$axis \cite{lattice}. These structural characteristics give rise to
the anisotropic physical properties observed in all the compounds in the $R$Sb$_{2}$ series \cite{Budko}.  Here, we have chosen to
focus on LaSb$_{2}$, since its low-temperature transport properties are not complicated by magnetic phase transitions which occur
in the other members of this series \cite{Budko}. 

	Single crystals of LaSb$_{2}$ were grown from high purity La and Sb by the metallic flux method \cite{Canfield}. 
The crystals grow as large flat layered plates which are malleable and easily cleaved.  The orthorhombic SmSb$_{2}$-structure type
was confirmed by single crystal X-ray diffraction.  Magnetization was measured with a commercial SQUID magnetometer, and transport
properties were performed using a standard 4-probe AC technique at 27 Hz at temperatures from 0.03-300 K and in magnetic fields up
to 45 T.  Hall effect measurements were made with a 5-wire geometry with data being taken in both positive and negative fields up
to 9 T.

	The in-plane zero-field electrical resistivity, $\rho_{ab}$, of single crystals of LaSb$_{2}$ was measured from 1.8-300 K and found
to be metallic ($d\rho/dT>0$).  The residual resistivity ratio (RRR) was large ($\rho_{ab}(300 K)/\rho_{ab}(2 K)\approx70-90$),
indicating a high sample quality.  The inset of Fig. 1 shows a plot of the in-plane resistivity versus $T^{2}$ at zero field.   The
data are linear, indicating that $\rho_{ab}\propto T^{2}$ up to 60 K.  This $T^2$ dependence has also been reported in YbSb$_2$ and
is believed to be a consequence of carrier-phonon scattering on a small cylindrical Fermi surface \cite{YbSb}.  The details of the
Fermi surface topology of LaSb$_2$ are not known, but it is indeed likely to be similar to that of YbSb$_2$ and hence the
$T^2$ behavior. Interestingly, LaSb$_2$ is a Type I superconductor with a Josephson coupling transition at $\sim0.35$ K
\cite{Young}.  In the main panel of Fig.\ 1 we show the in-plane transverse MR ($H\parallel c$) of LaSb$_{2}$ at 2 K.  The relative
MR is positive and nearly linear above 2 T with a high field slope of $\sim2$
$\mu\Omega$-cm/T.  The effect is also large, as the resistance increases by a factor of 90 from 0 to 45 T and shows no tendency
toward saturation. The linearity is unusual in that one expects the classical MR to saturate if the carriers are in closed Fermi
surface orbits or increase indefinitely as $H^2$ for open orbits \cite{Lifshitsbook}.  

	In order to characterize the sign and density of the charge carriers, we measured the in-plane Hall resistivity,
$\rho_{H}$, versus magnetic field for several different temperatures as shown in Fig.\ 2a.  The behavior in $\rho_{H}$ is
characteristically different from what would be expected for the classical Hall effect. The data are not linear in field
($\rho_{H}\propto H$) but are approximately quadratic, indicating a decreasing holelike carrier density. This effect is temperature
dependent, being largest at 2 K and progressively reduced with increasing temperature. There are two other significant features in
the data shown in Fig. 2a.  First, below 1 T the Hall resistivity at 2 K has a negative slope ($d\rho_{H}/dH<0$), and thus,
$\rho_{H}$ passes through zero with increasing field.  This is an indication of contributions from both electron and holelike
orbits at the Fermi surface, similar to what is observed in the low-temperature magnetotransport of YZn
\cite{Jan}. At fields much higher than the field at which
$\rho_{H}\sim 0$ (0.5 T), the Hall resistivity should be asymptotic to $\rho_{H}\sim HR_{h}R_{e}/(R_{h}+R_{e})$, where $R_h$ and
$R_e$ are the hole and electron Hall constants, respectively \cite{Ashcroft}.  Thus, in the high field limit $\rho_{H}$ is
dominated by the majority carriers which are clearly holes in LaSb$_2$. The second noteworthy feature in the data of Fig.\ 2a is
that the curves for the Hall resistivity all cross at $\sim 2$ T, indicating that the carrier density is independent of temperature
at this field.  We note that the Hall mobility, $\mu=\rho_H/(H\rho_{ab})$, is independent of field above $\sim1.5$ T and is
$\sim0.1 $ m$^{2}$/Vs below 20 K (Fig.\ 3 inset).  Thus, all of the Hall and transport data are in the regime
$\omega_c\tau\le5$, where $\omega_c$ and $\tau$ are the cyclotron frequency and relaxation time, respectively.  Finally, we have
demonstrated that the field dependence of the Hall resistivity depends uniquely on the component of the magnetic field
perpendicular to the $ab$-plane ($H_\perp$) by performing Hall measurements in tilted field.  The inset of Fig.\ 2a shows the Hall
resistivity at 2.5 K and 30 K plotted against $H_{\perp}$.  The data are essentially identical to those in the main panel of Fig.\
2a, indicating that the mechanism responsible for the anomalous Hall resistivity cannot be attributed to an orientation-independent
Zeeman splitting.  Likewise, the transverse MR with field oriented along the layers is an order of magnitude smaller \cite{Budko}.

	Further confirmation of the sign and field dependence of $n$ was obtained from the temperature
dependence of the thermoelectric power (TEP), $S$, as measured by the steady state method (Fig.\ 2b inset).  Like the Hall
coefficient, the sign of
$S$ is an indicator of the majority carrier sign, and at 300 K the TEP is negative and about an order of magnitude larger
than that of most good metals, such as Cu
\cite{TEP}.  The data are typical of many low-carrier-density metallic systems and suggest that LaSb$_{2}$ is $n$-type at all
temperatures in zero field.  The main panel of Fig.\ 2b shows the TEP plotted versus perpendicular field.  At 50 K the TEP is
constant in field, maintaining a small negative value ($\sim$ --4 $\mu$V/K) even up to 9 T.  The lower temperature data, however,
show a rather remarkable field dependence.  At 10 K and zero field, the TEP is $-$1.0 $\mu$V/K and then increases approximately
linearly in field with a positive  slope of about +1 $\mu$V/KT.  At $\sim 2$ T, the TEP passes through zero and then remains
positive at higher fields.  This increase in $S$ above 2 T is consistent with a depletion of the density-of-states with increasing
field \cite{TEP2}. 

	It is evident that the non-linear Hall resistance, the field dependent TEP, and the large, positive MR of Fig.\ 1 are related. 
Indeed, we have found that both the anomalous Hall resistivity and MR data can be modeled by assuming a single, field-dependent
carrier density that varies as $n\sim1/H$ at high field,

\begin{equation}
n(H)=\frac{n_{o}}{\sqrt{(1+(\frac{H}{H_{o}})^{2}}},
\end{equation}
where $n_{o}$ is the apparent carrier density at zero-field, and $H_{o}$ is a variable parameter which effectively determines a
characteristic field scale in the system.  The field-dependent Hall resistivity is,
$\rho_{H}(H)=H/en(H)$, where $e$ is the electron charge and $n(H)$ is given by Eq.(1).  The solid lines in the main panel and inset
of Fig.\ 2a are fits to the Hall resistivity data using this form, where $n_o$ and $H_o$ of Eq.(1) where varied for the best fit. 
Values of the variable parameter, $H_{o}$, extracted from the fits to $\rho_{H}$ are plotted as a function of temperature in the
inset of Fig.\ 4 (solid circles). Though $H_{o}$ saturates below 10 K, at higher temperatures it increases linearly with $T$.  The
solid line is a guide to the eye and has a slope 0.24 T/K.  This number is approximately equal to ${k_{B}}/{6\mu_{B}}$, where
$k_{B}$ and $\mu_{B}$ are the Boltzmann and Bohr magneton constants, respectively.  

 Like the Hall data, the MR of LaSb$_{2}$ can also be well described by the carrier density dependence of Eq.(1).  The
free-electron form of the  resistivity is simply, $\rho=1/e{\mu}n(H)$, where $\mu$ is the carrier mobility.  Since the mobility
above 1.5 T is only weakly field dependent, the relative MR is well described by $\rho(H)/{\rho(0)}=n_o/n(H)$.  The  solid line in
Fig.\ 1 is a least-squares fit to the MR where only $H_o$ in Eq.(1) was varied.  The fit captures most of the qualitative behavior
of data, including the linear dependence at high field.  The $H_o$ values extracted from fits at different temperatures are shown
as open circles in the inset of Fig.\ 4.  Though smaller than the values obtained from the Hall data, they have the same high
temperature slope, ${k_{B}}/{6\mu_{B}}$.   

	Figure 3 shows the temperature dependence of the apparent Hall carrier density, $n_{app}=H/(e\rho_H)$, at several magnetic
fields.  At low temperature $n_{app}$ is very sensitive to field, with the carrier density saturating below 10 K at all fields.  At
2 K, for example, the carrier density at 1 T is reduced by an order of magnitude in a field of 9 T.  At 2 T, which is the
cross-over field in both the Hall resistivity and TEP, $n_{app}$ is temperature independent.  The resistivity of LaSb$_{2}$ at 2
T is also temperature independent below 30 K.  For fields greater than 2 T, $n_{app}$ decreases with increasing temperature, and
for fields less than 2 T, $n_{app}$ {\em increases} with increasing temperature.  Interestingly, the data in Fig.\ 3 are reminiscent
of $\rho$ vs. $T$ plots of systems displaying a magnetic field induced metal-insulator or superconductor-insulator transition
\cite{SIT}.  Indeed, the 2-T curve may represent a separatrix between two low-temperature phases of the system \cite{Fisher}. 
Clearly the high-field phase appears to be somewhat semiconducting in character in that the carrier density decreases with
decreasing temperature.  Nevertheless, $\rho_{ab}$ vs. $T$ measurements indicated the system was still metallic at 45 T.

	While the results from the magnetotransport measurements outlined above indicate a reduction of the carrier density with
increasing  field, more compelling evidence to this fact is provided by a thermodynamic measurement of the bulk susceptibility
of LaSb$_{2}$. The magnetization was observed to be diamagnetic and slightly super-linear with the magnetic field applied along
the $c$-axis \cite{Budko}.  In the analysis that follows we assume that the magnetization arises from two primary contributions: 
Larmor diamagnetism (of the closed-shell ion cores) and Pauli/Landau paramagnetism of the conduction electrons. By subtracting the
purely linear Larmor background from the bulk magnetization data, we were able to isolate the Pauli and Landau contributions which
are both proportional to the density of states at the Fermi level, $D(\epsilon_{F})$.  The main panel of Fig. 4 shows
$\chi-\chi_{L}$ versus field, where $\chi$ and $\chi_{L}$ are the bulk and Larmor susceptibilities, respectively.  For a first
approximation, the Larmor term  was set equal to three times the molar susceptibility of a xenon ion \cite{Ashcroft}, i.e.
$\chi_{L}=3$ x (--43 x 10$^{-6}$ cm$^{3}$/mole).  The data points in Fig. 4 now represent an estimation of the Pauli and Landau
contributions to the susceptibility.  For a free-electron gas, $D(\epsilon_{F})\propto n^{1/3}$, so that using Eq.(1) we have,

\begin{equation}
\chi - \chi_{L}={\alpha}{\Bigl[ 1+\Bigl(\frac{H}{H_o}\Bigr)^2\Bigr]^{-\frac{1}{6}}}.
\end{equation}
The solid line in Fig.\ 4 is a fit to the susceptibility using Eq.(2), where
$\alpha$ and $H_o$ were varied for the best fit. Even considering simplifications, such as neglecting
band effects (i.e. effective masses), electron-electron interactions, and uncertainties associated with assessing the diamagnetic
background, the overall structure of the fit is in good agreement with the susceptibility data and is consistent with the
non-linear Hall data of Fig.\ 2.    

	In conclusion, we find that the linear MR of LaSb$_2$ is due to an orbital magnetic field suppression of the density-of-states
with a corresponding attenuation of the holelike carrier density that does not saturate up to 45 T.  Above 2 T, where the MR
becomes linear, we find that the temperature derivative of the carrier density changes from negative to positive, suggesting that
the system may undergo a metal-insulator transition at sufficiently high field. A large, linear MR has also been reported in a
number of the magnetic rare-earth diantimonides \cite{Budko}, which have the same structure as LaSb$_2$.  It has been conjectured
that the MR of these materials is a quantum manifestation of the peculiar Fermi surface topology \cite{Abrikosov2} distinct from
a field dependent $n$.  Magnetotransport and de Hass-van Alphen studies under hydrostatic pressure should prove
interesting and may help to disentangle electronic structure effects from possible field-modulated localization \cite{Phillips} in
these highly anisotropic metals.

	We are grateful to D. Browne and R. Kurtz for many enlightening discussions.  This work was done, in part, at the National High
Magnetic Field Laboratory and was supported by the NSF under Grants DMR 99-72151 and DMR 01-03892.  We also acknowledge support of
the Louisiana Education Quality Support Fund under Grant No. 2001-04-RD-A-11.


%
\newpage
\begin{figure}
\vspace{1.5in}
\centerline{\epsfig{file=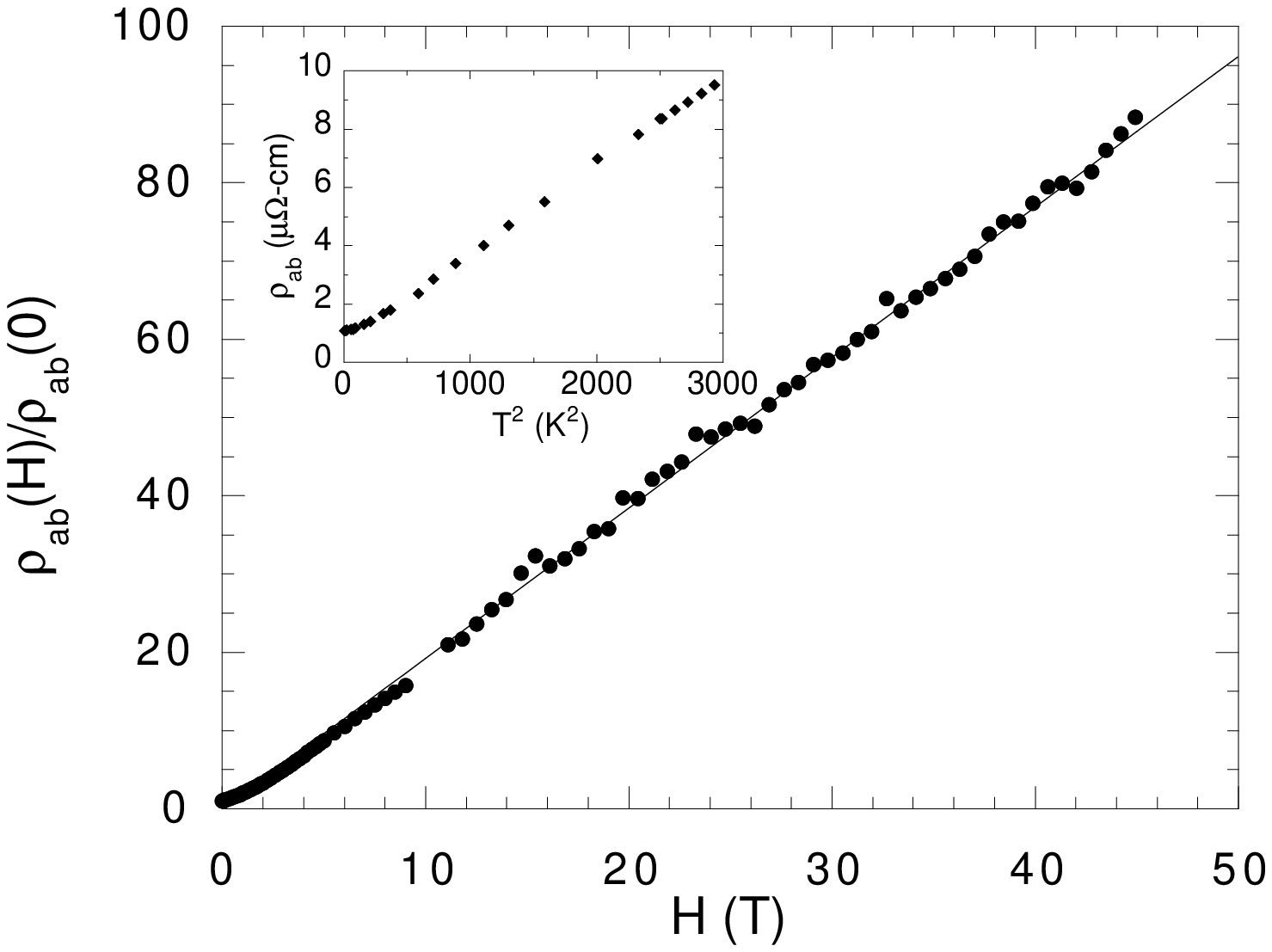,width=6.0in}}
\vspace{0.6in}
\caption{In-plane resistivity of LaSb$_2$ at 2 K as a function of perpendicular magnetic field, $\rho_{ab}(0)=1.0$ $\mu\Omega$-cm. 
The solid line is a fit to the data using Eq.(1) as described in the text. Inset: $T^2$ temperature dependence in zero field.}
 \label{Figure 1}
 \end{figure}
\newpage

\begin{figure}
\vspace{0.2in}
\centerline{\epsfig{file=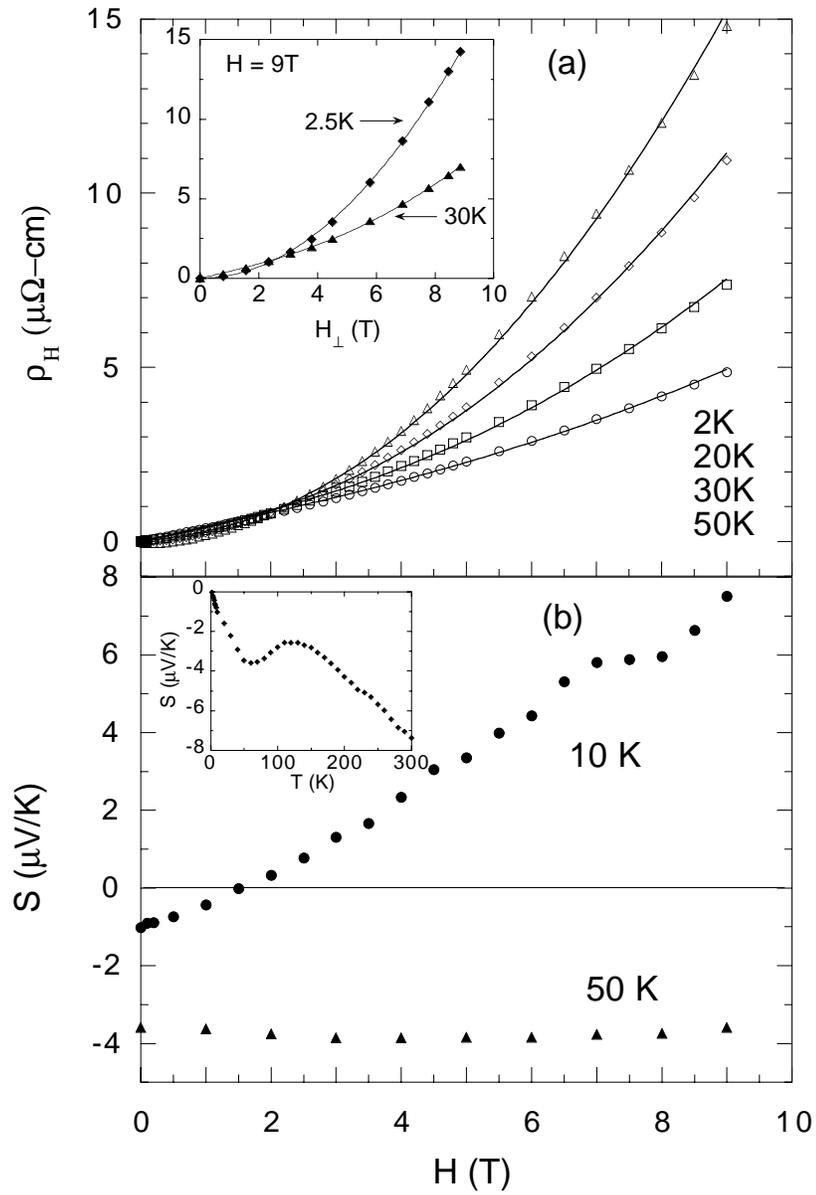,width=5.0in}}
\vspace{0.1in}
\caption{(a).  Hall resistivity as a function of magnetic field oriented perpendicular to the layers.  Inset: Hall resistivity
obtained by rotation in constant magnetic field, where $H_{\perp}$ is the field component perpendicular to the layers. The solid
lines are fits to the data using Eq.(1) as described in the text. (b) Thermoelectric power as a function of perpendicular magnetic
field. Note the sign change at $\sim2$ T in the 10 K data. Inset: Temperature dependence of the thermoelectric power in zero
field.}
 \label{Figure 2}
 \end{figure}
\newpage

\begin{figure}
 \vspace{02.5in}
\centerline{\epsfig{file=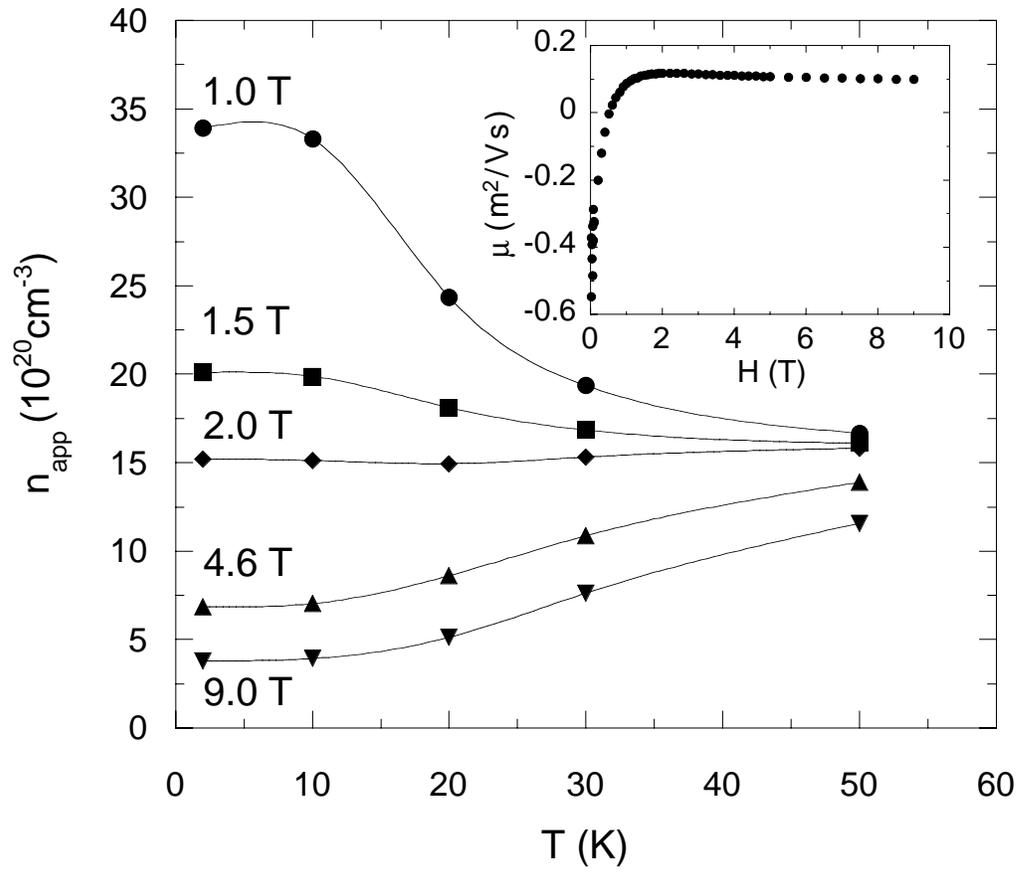,width=6.0in}}
\vspace{0.6in}
\caption{Temperature dependence of the apparent carrier density as derived from Hall data in Fig.\ 2. 
Note that the 2-T carrier density is independent of temperature.  The lines are provided as a guide to the eye.  Inset: Field
dependence of the apparent mobility at 2 K.  The large negative values of $\mu$ are an artifact of $\rho_H\le0$ below 1T.}
 \label{Figure 3}
 \end{figure}
\newpage

\begin{figure}
 \vspace{2.5in}
\centerline{\epsfig{file=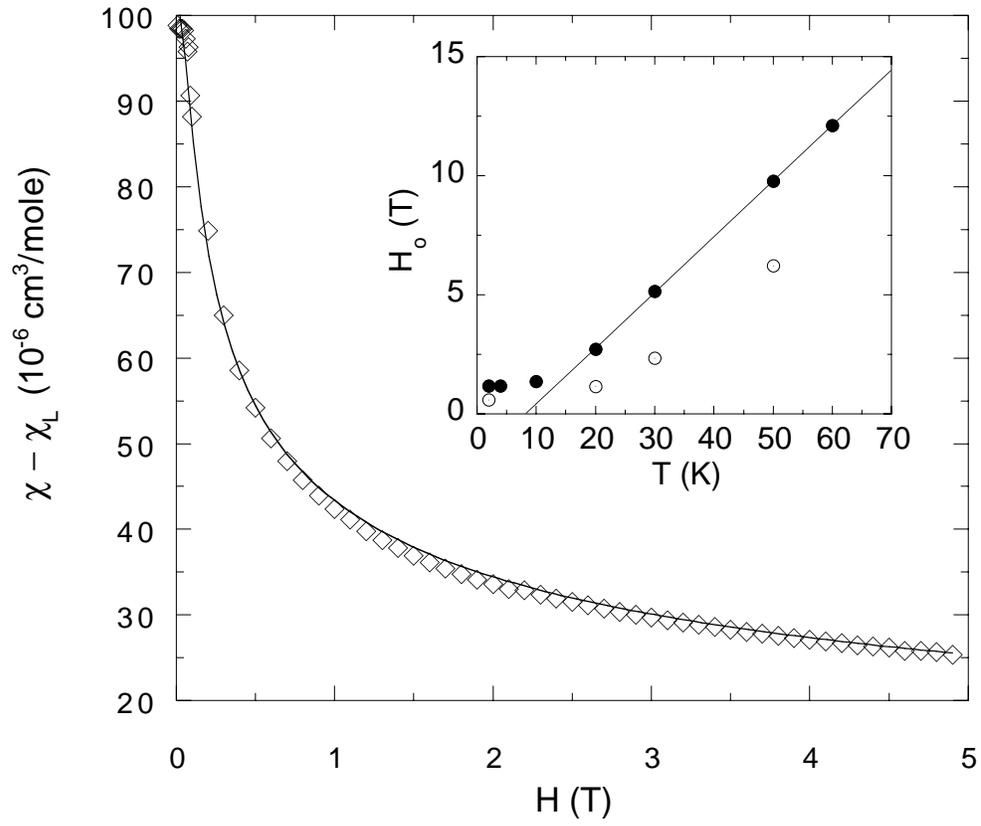,width=6.0in}}
\vspace{0.3in}
\caption{Susceptibility at 2 K as a function of perpendicular magnetic field.  The Larmor susceptibility has been subtracted off
leaving behind the Pauli and Landau contributions which are both proportional to the density of states. The solid line is a fit to
the data using Eq.(2). Inset: Solid symbols are $H_o$ values obtained from fits to the data in Fig.\ 2, and the open symbols are
values obtained from fits to the relative MR.  The solid line is provided as a guide to the eye and has a slope of
${k_{B}}/{6\mu_{B}}$.}
 \label{Figure 4}
 \end{figure}
\newpage

%

%


\begin{references}
\bibitem{lattice}X-ray analysis determined the lattice constants to be 0.6319 nm, 0.6173 nm,
and 1.857 nm for the $a$, $b$, and $c$ crystallographic directions, respectively.
\bibitem{TEP2}$S$ is proportional to the logarithmic derivative of the density-of-states, see Ref.\cite{TEP}.
\end{references}
\end{document}